\documentstyle[aps,prb,epsf,psfig]{revtex}
\begin{document}

\twocolumn[\hsize\textwidth\columnwidth\hsize\csname@twocolumnfalse\endcsname

\title{Optimal basis set for electronic structure calculations in 
       periodic systems}
\author{Sandro Scandolo$^{a,b}$ and Jorge Kohanoff$^{b,c}$}
\address{
         $a)$ International School for Advanced Studies (SISSA) and INFM,
              Via Beirut 4, I-34014  Trieste, Italy\\
         $b)$ International Centre for Theoretical Physics (ICTP),
              I-34014 Trieste, Italy\\
         $c)$ Atomistic Simulation Group, The Queen's University,
              Belfast BT7 1NN, Northern Ireland \\
        }
\maketitle

\begin{abstract}
An efficient method for calculating the electronic structure of
systems that need a very fine sampling of the Brillouin zone is presented.
The method is based on the variational optimization of a {\em single} 
(i.e. common to all points in the Brillouin zone) basis 
set for the expansion of the electronic orbitals. Considerations from $k\cdot 
p$-approximation theory help to understand the efficiency of the method. 
The accuracy and the convergence properties of the method as a function
of the optimal basis set size are analysed for a test calculation on
a 16-atom Na supercell. 
\end{abstract}

\date{\today }

\vspace{0.5cm}
]

\section{Introduction}

First-principles methods have become a widespread tool for studying the 
microscopic aspects of a wide class of condensed systems, including their
crystalline phases. No empirical modelling of the interatomic interactions
is required in first-principles simulations, since the electronic ground 
state is evaluated at each atomic configuration\cite{carpar}. 
In a large set of systems, mostly in metals, but also in some molecular 
crystals like hydrogen\cite{mazin}, a very fine sampling of the 
Brillouin Zone (BZ) is required to provide a sufficiently accurate 
description of the electronic ground state. Since the computational load
for refining BZ sums grows linearly with the number of sampling points 
(k-points), 
special k-point grids that minimize sampling errors, as well as interpolation
techniques (the so called tetrahedron method\cite{taut}) have been introduced
in the past\cite{balde,cha-co,mp}.  
However, little attention has been paid 
to the fact that Bloch states at nearby k-points are not totally uncorrelated.
In fact, once the Bloch states at a given k-point are known, those at a 
closeby k-point can be easily estimated by perturbation theory, 
within the so-called $k\cdot p$ approximation\cite{harrison}.
This extremely helpful piece of 
information has been little exploited so far in standard first-principles 
approaches, where for each k-point a full solution of the Kohn-Sham 
equations is instead calculated independently, e.g. through diagonalization
of the Kohn-Sham hamiltonian on a finite basis set (typically plane waves
in the case of crystalline systems).  
An approximate technique based on the 
exact solution of the Kohn-Sham hamiltonian at a few k-points has
been proposed\cite{payne1}, but the analysis of the errors
associated with such method, as well as their correction, has proved 
rather difficult\cite{payne2}.

In this work we propose a simple and efficient method for solving the Kohn-Sham 
equations for a very large number of k-points in the BZ, with an effort
that in most cases is comparable to that of solving the problem for just
a few k-points. The method is based on a rigorous formulation of the ideas
sketched above, and relies on the construction of a single, optimal basis set 
for the expansion of the Bloch orbitals, which is common to all k-points.
The application of this method to the study of the compressed phases of 
molecular hydrogen has been recently reported \cite{ksct,mrs,ksdt}. 
In Section II we formulate the method.
In Section III we analyse the reliability of the method on a test case 
(16 Na atoms in the bcc lattice), and in Section IV we analyse the 
computational cost. Section V contains the conclusions.

\section{The method}

In first-principles simulations
the electronic ground state is calculated within the Kohn-Sham, self-consistent
one-electron formalism of density functional theory \cite{kohn-sham}.
Translational invariance is introduced
by averaging the electronic density and the total energy of the system on a
finite number of k-points in the BZ:
\begin{equation}
\rho ({\bf r})=\sum_{{\bf k}}^{BZ}\omega _{\bf k}~\sum_i~f_i({\bf k})\mid
\psi _i^{{\bf k}}({\bf r})\mid ^2~~,  \label{rho}
\end{equation}
\begin{equation}
E[\rho]= T+E_{{\rm ext}}[\rho ]+E_{{\rm H}}[\rho ]+E{\rm xc}[\rho ]
\label{energy}
\end{equation}
with
\begin{equation}
T = 
\sum_{{\bf k}}^{BZ} \omega _{\bf k}~\sum_i~f_i({\bf k})\int \psi _i^{*~%
{\bf k}}({\bf r})\left( -\frac{\nabla ^2}2\right) \psi _i^{{\bf k}}({\bf r}%
)~d{\bf r}~~,
\label{kin}
\end{equation}
where $i$ labels bands, {\bf k} a k-vector in the BZ, $\omega _{\bf k}$
its weight,
$f_i({\bf k})$ are occupation numbers, and $\psi _i^{{\bf k}}({\bf r})$ 
are the Kohn-Sham orbitals. The last three terms in (\ref{energy}) are the 
external, Hartree, and exchange-correlation contributions to the
energy, and depend on the BZ averaged electronic density (\ref{rho}). 
Atomic units are used throughout this work.

Variational equations for the periodic part of the Kohn-Sham orbitals
$u_i^{\bf k}({\bf r}) = \exp{(-i{\bf k\cdot r})}
 \psi _i^{\bf k}({\bf r})$ 
are readily obtained from expression~(\ref{energy}):
\begin{equation}
\left(-\frac{\nabla^2}{2}-i{\bf k}\cdot\nabla+\frac{k^2}{2}
      +V[\rho({\bf r})]\right)
u_i^{{\bf k}}({\bf r})= \epsilon_i^{{\bf k}} u_i^{{\bf k}}({\bf r})  ~~,
\label{kohnsham}
\end{equation}
where $V[\rho]$ is the Kohn-Sham potential, 
containing external, Hartree and exchange-correlation contributions. 
Normally, equations (\ref{kohnsham}) are solved independently for each
k-point, a new density is constructed via (\ref{rho}), 
and the cycle is iterated to self-consistency. 
Equations (\ref{kohnsham}) are often solved by expanding the orbitals 
corresponding to the various {\bf k}-points in a basis set of plane waves, which 
remains unchanged when the atomic configuration is changed. 
The choice of a plane-wave basis set 
offers some practical advantages, including the convenience of calculating 
a number of quantities (matrix elements, form factors, etc) a single time 
throughout the calculation. However, working with a plane-wave basis set 
becomes particularly demanding from the computational point of view when a 
large supercell and/or a large number of $k$-points has to be considered, a 
situation which is often encountered when dealing with metals. 
Moreover, a plane-wave basis set cannot straightforwardly 
exploit the similarity of Bloch states at closeby k-points. 

In the following we will show that relaxing the requirement of a ``simple''
but inmutable basis set
in favor of a variational search for the ``best'' basis set at each atomic 
configuration may improve substantially the efficiency of BZ summations. 
To this aim, let us expand the Kohn-Sham orbitals on a generic basis set 
$\phi_i ~(i=1,...,N)$ of size $N$: 
\begin{equation}
u_i^{{\bf k}}({\bf r})=\sum_{j=1}^N~a_{ij}({\bf k})~\phi_j({\bf r})~~.
\label{aij}
\end{equation}
In the case of standard calculations $\phi_i$ would be a plane wave
-- $\phi_i({\bf r}) \sim \exp(i{\bf g}\cdot{\bf r})$, being {\bf g} 
a reciprocal lattice vector -- and solving the 
Kohn-Sham equations would amount to minimize the energy
functional (\ref{energy}) with respect to the expansion coefficients 
$a_{ij}$, without modifying the basis set. 
In the present approach we minimize the energy
functional (\ref{energy}) with respect to {\em both} the expansion coefficients
$\{ a_{ij}\}$ {\em and}~the basis set functions $\{ \phi_i \}$. 
In other words, we look simultaneously
for the ground state orbitals {\em and} for basis set that better 
describes them. For the sake of clarity, we note that this would be a
redundant request for calculations with a single k-point, where a single 
set of Kohn-Sham equations have to be solved. The request of an optimal basis
set becomes non redundant when the basis set is used to describe
Kohn-Sham equations at {\em many} k-points.

Explicit minimization of (\ref{energy}) with respect to,
simultaneously, $a_{ij}$ and $\phi_i$ leads to the following system of
equations. For the expansion coefficients we have the usual eigenvalue 
equations:
\begin{equation}
\sum_{j=1}^N\left( \lambda _{ij}^0+\frac{k^2}{2}\delta _{ij}+{\bf k}\cdot
{\bf p}_{ij}\right) a_{jl}({\bf k})=\epsilon _i^{{\bf k}}a_{il}({\bf k})~~,
\label{matrix}
\end{equation}
with
\begin{equation}
\lambda _{ij}^0=\left\langle \phi_i\left| \left( -\frac{\nabla ^2}2+V_{%
{\rm KS}}[\rho ]\right) \right| \phi _j\right\rangle
\end{equation}
and
\begin{equation}
{\bf p}_{ij}=-i\left\langle \phi _i\left| {\bf \nabla }\right| \phi
_j\right\rangle ~~.
\end{equation}
For the basis-set functions we have
\begin{eqnarray}
\frac{\delta E[\phi ,{\bf a}]}{\delta \phi _i^{*}({\bf r})}&=&\sum_{j=1}^N
B_{ij}^{(0)}\left[-\frac{\nabla^2}{2}+v_{\rm KS}[\rho]\right]~
\phi _j^{*}({\bf r})+ \nonumber \\
&+& \sum_{j=1}^N \left\{-i{\bf B}_{ij}^{(1)}\cdot {\bf \nabla}+
B_{i}^{(2)}\delta_{ij}\right\}~\phi _j^{*}({\bf r})+ \nonumber \\
&+& \frac{\delta E_{NL}[\phi ,{\bf b}]}{\delta \phi _i^{*}({\bf r})}
  = \sum_{l=1}^N \Lambda_{il} \phi_l({\bf r}) ~~,
\label{forbkdp}
\end{eqnarray}
with
\begin{eqnarray}
  B_{ij}^{(0)}&=&\sum_{\bf k}\omega_{\bf k}~b_{ij}({\bf k}) \nonumber \\
  {\bf B}_{ij}^{(1)}&=&\sum_{\bf k}\omega_{\bf k}~{\bf k}~b_{ij}({\bf k}) \\
  B_{i}^{(2)}&=&\frac{1}{2}\sum_{\bf k}\omega_{\bf k}~k^2~b_{ii}({\bf k})
  \qquad  \nonumber
\end{eqnarray}
and
\begin{equation}
  b_{jl}({\bf k})=\sum_{i=1}^{N}f_i({\bf k})~a_{ij}^{*}({\bf k})
  a_{il}({\bf k}) ~~.  
\label{bdkp}
\end{equation}
The $\Lambda$ coefficients in (\ref{forbkdp}) are Lagrange multipliers 
introduced to enforce the orthonomalization of the basis functions 
\cite{carpar}.
The electron density in (\ref{forbkdp}) is constructed from the Kohn-Sham
orbitals (\ref{aij}) using (\ref{bdkp}), as
\begin{equation}
\rho ({\bf r})=\sum_{j=1}^N\sum_{l=1}^N\phi _j^{*}({\bf r})~\beta _{jl}
~\phi_l({\bf r})~~,  \label{rhokdp}
\end{equation}
where
\begin{equation}
\beta _{jl}=\sum_{{\bf k}}\omega _{{\bf k}}~b_{jl}({\bf k}) ~~.
\label{betakdp}
\end{equation}
More detailed expressions for the above equations as well as for atomic
forces and stress are given in the Appendix, for the case of pseudopotential
plane wave (PPW) calculations.

The problem of simultaneously solving Kohn-Sham equations for $N_k$
k-points independently, as done in standard electronic structure 
calculations, is here replaced by the problem of solving a {\em single} 
set of Kohn-Sham-like equations, eqs. (\ref{forbkdp}), supplemented with 
$N_k$ diagonalizations of the $N\times N$ matrix in (\ref{matrix}).
As we will show in the next Section, the number of basis functions $N$ 
required for the optimal basis set $\{ \phi_i\}$ can be chosen to be much 
smaller than that in the basis set used to describe each single $\phi _i$,
e. g. the number of plane wave components in PPW calculations. This implies
that, if $N_k$ is large, the computational effort required 
by this procedure is much smaller than the effort required to solve 
the standard set of Kohn-Sham equations at all k-points. The main
limitation of the method resides, then, in the size $N$ of the optimal 
basis set. If $N$ turns out to be comparable to the original number of
basis functions, then there is no point on pursuing this strategy.

While in standard
calculations the number of Kohn-Sham equations is roughly (exactly
for insulators) half the number of electrons, in our approach $N$ 
has to be sufficiently large to guarantee the proper description of 
the Kohn-Sham orbitals (\ref{aij}) at all k-points. A full analysis 
of the convergence of the method as a function of $N$ will be presented
later, for a specific case. 
There are however reasons to expect that $N$ need not be too large.
In fact, a reasonable approximation to the {\em exact} optimal basis set 
is given by the Kohn-Sham orbitals at one selected k-point, say at
zone center ($\Gamma$). We call this basis set: $\{\phi_i^\Gamma\}$. 
Since the optimal basis set is defined as the best basis set in a 
variational fashion,$\{\phi_i^\Gamma\}$ will necessarily
provide a worse description of the Kohn-Sham orbitals (except, of course, 
at $\Gamma$) than $\{\phi_i\}$. The convergence properties of
$\{\phi_i\}$ will thus be better than those of $\{\phi_i^\Gamma\}$.
If the BZ of the system of interest is sufficiently small (which is 
equivalent to ask that the unit cell of the calculation is large),
then $k\cdot p$ perturbation theory suggests that Kohn-Sham orbitals
at any k-point other than $\Gamma$ may be expressed in terms of a 
few number of energetically close $\Gamma$-point orbitals $\{\phi_i^\Gamma\}$
\cite{harrison}.
In other words, only a few excited $\Gamma$-point orbitals are required 
for a reasonable description of the orbitals at all other k-points.
This property has been already noted and used in the context of 
first-principles electronic structure calculations\cite{payne1,payne2,payne3}. 
Our approach differs from that of Robertson and Payne \cite{payne1,payne2} 
in the fact that their basis set is given by the Kohn-Sham orbitals 
at $\Gamma$ (or at a small set of k-points), while our method includes a 
variational search for the best basis set. A comparison of the convergence 
properties of the two approaches is given in the next Section.

We also observe that our method has only one variational parameter,
namely the size $N$ of the basis set. This implies that systematic 
improvements of the accuracy can be achieved by simply increasing $N$.

Finally, we notice that eqs. (\ref{forbkdp}), like ordinary Kohn-Sham 
equations, have to be solved self-consistently, since the Kohn-Sham 
potential $v_{KS}$ 
depends on the density, and thus on wavefunctions. However, unlike ordinary 
Kohn-Sham equations, eqs. (\ref{forbkdp}) depend also on the expansion 
coefficients 
$a_{ij}$. The ordinary self-consistent iterative procedure has, consequently,
been modified in the following way: 

\begin{enumerate}
\item{an initial guess of the density is provided, e.g. as a sum of atomic 
densities; the $a_{ij}$ coefficients are initially set to $\delta_{ij}$,}
\item{eqs. (\ref{forbkdp}) are solved,}
\item{the orbitals (which are no longer Kohn-Sham orbitals) are used to 
construct $\lambda^0_{ij}$ and ${\bf p}_{ij}$,}
\item{eq. (\ref{matrix}) is solved for each k-point, via direct 
diagonalization,}
\item{a new density is constructed,}
\item{back to 2.}
\end{enumerate}

Notice that this procedure does not involve a major implementation effort,
but only a few modifications to the usual self-consistency methods.

\section{Convergence of the optimal basis set}

In order to analyse the convergence properties of the method 
we have chosen the case of Na metal, which is properly described with
a soft, local pseudopotential\cite{top-hop}. 
The Ceperley-Alder local density functional was adopted 
for the exchange-correlation term~\cite{cep-ald}.
Sixteen Na atoms in the bcc structure have been placed in a simple 
cubic supercell ($2\times 2 \times 2$ bcc conventional cells), 
and the full Brillouin zone of 
the supercell has been sampled with 108 k-points\cite{mp}. 
A plane wave energy cutoff of 15 Ry was chosen, amounting to 1596 
plane waves in the expansion of the optimal basis set. 
The convergence of the method has been analysed by performing calculations
with $N=19$, 32, 51, 81, and 1596~\cite{shells}. 
Since our aim is at checking the convergence properties of the method, 
and not the overall accuracy as determined by the other approximations 
(LDA, k-point grid, etc.), we will consider as converged the values 
obtained with $N= 1596$ and $N_k=108$. As a comparison, we have also 
determined the 
convergence properties of a calculation where the basis set, instead of
being ``optimal'', is chosen to be the set of Kohn-Sham orbitals resulting 
from a self-consistent calculation using the $\Gamma$-point only. This
correponds to the approach of Ref. \cite{payne1,payne2}.

\vspace{0.5truecm}
\begin{figure}[thb]
\centerline{\epsfxsize=3.0truein
\psfig{figure={ene_ts.eps},angle=270,width=6.5cm,height=5.5cm}}
\vspace{0.3truecm}
\caption{Convergence of the total energy of solid $bcc$ Na using a 16-atom
supercell, as a function of the number of the basis set size $N$. Dashed
lines correspond to the approach of Ref. \cite{payne1,payne2}, and solid
lines to the present method. The horizontal line is the converged value.}
\end{figure}

In Figs. 1 to 3 we show the convergence of the different physical properties
as a function of $N$, namely the total energy, stress, and electronic density
of states. The total energy (Fig. 1) shows an extremely good convergence 
even for the smallest value of $N$. The accuracy is better than 0.5 \%
for $N=19$, and better than 0.1 \% for $N=32$ states. The same behavior is 
observed in the force constant calculated for a distortion along an optical 
phonon: the force is already converged to better than 0.1 \% 
using 32 states in the expansion. Thus, the errors introduced in energies
and forces due to truncation of the expansion are much smaller than, e.g., the
typical accuracy of the usual density functionals used, namely the LDA or
the GGA. Stresses also show good convergence properties, as reported in Fig. 2. 
Variations of stress ($\Delta\sigma$) with respect to $N$ should in this
case be compared with the value of the bulk modulus of Na ($B=64$ kbar).   
In fact, an error $\Delta\sigma$ in the stress yields an error in the 
equilibrium lattice spacing that can be estimated as 
$\Delta a /a \sim \Delta\sigma /3B$. Thus, even with $N=19$ the equilibrium 
lattice spacing is converged within less than 1 \% . 

\vspace{0.7truecm}
\begin{figure}[thb]
\centerline{\epsfxsize=3.0truein
\psfig{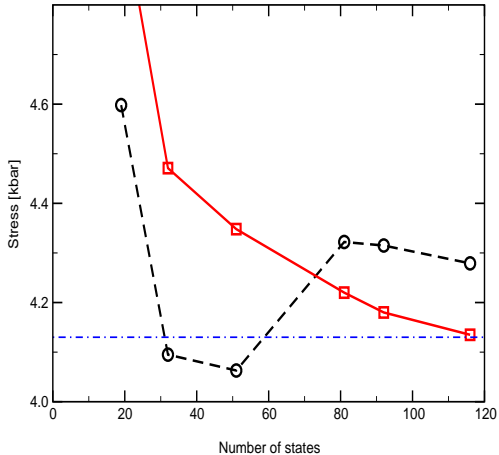}}
\vspace{0.5truecm}
\caption{Convergence of the trace of the stress tensor of solid $bcc$ Na
using a 16-atom supercell, as a function of the basis set size $N$. Dashed
lines correspond to the approach of Ref. \cite{payne1,payne2}, and solid
lines to the present method. The horizontal line is the converged value.}
\end{figure}

The electronic density of states (EDOS) is also sufficiently accurate for the 
relevant portion of the spectrum, namely the one below the Fermi energy 
($E_F$), using $N=32$. In contrast, $N=19$ appears to give an inaccurate EDOS, 
particularly in the vicinity of $E_F$, as expected in a $k\cdot p$ picture.
In fact, as discussed in Section II, a small basis set is more likely to
affect the accuracy in the description of states with higher energy, i.e.
closer to $E_F$, due to the insufficient number of high energy states in
the optimal basis set.

\vspace{0.7truecm}
\begin{figure}[thb]
\centerline{\epsfxsize=2.7truein
\epsffile{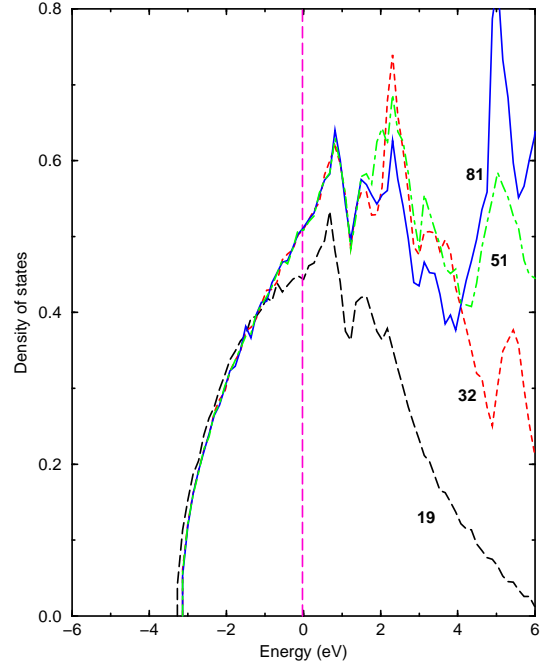}}
\vspace{0.5truecm}
\caption{Convergence of the electronic density of states of bcc Na using a
16-atom supercell, as a function of the basis set size $N$.
The different curves correspond to different values of $N$: long-dashed
($N=19$), short-dashed ($N=32$), dot-dashed ($N=51$), and solid ($N=81$).
Only results obtained with the present method are reported.}
\end{figure}

We now compare the results obtained with our method against those obtained
with the method of Ref. \cite{payne1,payne2}. 
Our energies are always lower, for a given basis set size $N$, 
than energies calculated using the method of Ref. \cite{payne1,payne2}. 
This is not unexpected, since the optimal basis set is variationally 
the best basis set. Moreover, we find that the method of 
Ref. \cite{payne1,payne2} provides results with an accuracy similar to ours
only when the size of the basis set is 30-50 \% larger than ours.   The most
striking difference, however, is in the convergence of the stress tensor.
While our method converges very smoothly to the exact value, the method 
of Ref. \cite{payne1,payne2} shows an erratic behavior, that clearly prevents
every attempt to determine the exact value on the basis of size scaling. 
We believe that such a marked difference in the convergence properties 
has to be traced to the variational nature of our approach.
Therefore, our method not only improves on that of Ref. \cite{payne1,payne2}
in the rate of convergence (which, in the case of energy, results in a 
30-50 \% saving in computational time, as shown in the next Section), 
but shows a much smoother behavior as
a function of $N$, resulting in a more accurate estimate of the error 
introduced by the finite size of the basis set.

\section{Estimate of the computational load}

The computational effort in a standard plane wave calculation scales as
$N_k N_{bands} N_{pw}^2$, where $N_{bands}$ is the number of occupied bands
(equal or slightly larger than half the 
number electrons) and $N_{pw}$ is the size of the plane 
wave basis set. This estimate is based on the assumption that standard 
diagonalization methods are used, such as iterative diagonalization, where
the effort of determining the first $N_{bands}$ eigenvalues of a 
$N_{pw}\times N_{pw}$ matrix scales like $N_{bands} N_{pw}^2$. Since every
k-point requires a separate diagonalization, the total effort scales as
$N_k N_{bands} N_{pw}^2$.

In the present approach, if we assume that the effort of diagonalizing 
$N_k$ times the $N\times N$ matrix (\ref{matrix}) is negligible with respect
to the effort required to solve eqs. (\ref{forbkdp}), then the load scales
as $N N_{pw}^2$. Thus, a gain of a factor $N_k N_{bands} / N$ is expected 
with our method. Since in typical applications $N \sim 2\div 4~N_{bands}$,
and $N_k \sim 10^2$, a computational gain of more than one order of
magnitude is obtained. The neglected term becomes increasingly important 
as the size of the system ($N$) increases, scaling as $N_k\times N^3$.
Therefore, it becomes dominant for sufficiently large $N$, thus 
discouraging the use of the present methodology in the case of large unit 
cells and BZs sampled in a few k-points.

The above estimate is based on the assumption that solving eqs. (\ref{forbkdp})
is as time consuming as solving the standard Kohn-Sham equations,
except for the fact that the requested number of states is larger.
Our experience on Na (this work) and molecular hydrogen \cite{ksct,mrs,ksdt} 
shows that the computational overload of solving eqs. (\ref{forbkdp}) 
basically scales linearly with the number of states; the extra effort
required for solving a single state is negligible when compared to 
that of solving one of the usual Kohn-Sham states. Nevertheless, we 
are currently not in a position to generalize this conclusion to a generic 
system.

\section{Summary and conclusions}

We have presented and analysed the convergence properties of a method
for performing electronic structure calculations and first-principles
molecular dynamics simulations for systems that need a very fine sampling
of the electronic Brillouin zone. The method is based on the construction
of a basis set which optimally describes the Kohn-Sham orbitals
at all k-points. The rapid convergence of the method on the size $N$ 
of the basis set is connected with the properties of 
the $k\cdot p$ approximation and has been
fully analysed in the case of a sixteen Na atoms supercell.
We have shown that $N$ can be kept reasonably small, 
of the order of a few times the number of occupied bands (2 to 4 
times, depending on the accuracy desired), resulting in a 
gain in computational load of about one order of magnitude, with
respect to standard methods. 

Even if our analysis was mainly done in connection with PPW calculations, 
the same philosophy can be adopted in other electronic structure schemes
like the LAPW method \cite{lapw}, provided that the number of orbitals
required in the optimal basis set remains significantly smaller than 
the number of basis functions in the original basis set.

In conclusion, this technique constitutes a useful tool for studying 
the electronic, structural and dynamical properties of metals 
and more generally for any system whose proper
description requires a fine sampling of the BZ, like, 
for example, molecular hydrogen.

\section{Acknowledgements}

SS acknowledges partial support from INFM through ``Iniziativa Calcolo 
Parallelo'', and from MURST through COFIN 1999. JK ackowledges the
hospitality of the Universidad Nacional de San Luis, Argentina, where
part of this work was carried out.

\section{Appendix} 

In this appendix we give some useful expressions for the total energy,
atomic forces and stress, in the case of pseudopotential plane wave calculations. 

The different contributions to the total energy (\ref{energy}) can be 
divided into the class of terms that only depend on the density (the
hartree, the exchange-correlation, and the local part of the pseudopotential),
and the class of terms that explicitly depend on the wavefunctions
(the kinetic and the nonlocal pseudopotential contributions). While the
former can be calculated in the usual way once the density is known 
(see below for an efficient way to evaluate density), the latter can 
be expressed more conveniently as follows. For the kinetic energy we have
\begin{eqnarray}
T&=&\sum_{{\bf k}}\omega_{{\bf k}}\sum_{i=1}^N\sum_{j=1}^N
b_{ij}^{{\bf k}}~\left\{t_{ij}^0[\phi]+{\bf k}\cdot {\bf p}_{ij}+\frac{k^2}{2}
\delta _{ij}\right\}=\nonumber \\
&=&\sum_{i=1}^N~\left\{B_{ii}^{(0)}t_{ii}^0[\phi]+B_{i}^{(2)}\right\}+
\nonumber \\
&+& \sum_{i=1}^N\sum_{j\ne i}^N~\left\{B_{ij}^{(0)}t_{ij}^0[\phi]-
i{\bf B}_{ij}^{(1)}\cdot {\bf p}_{ij}\right\} ~~,
\end{eqnarray}
with
\begin{equation}
t_{ij}^0[\phi]=<\phi _i\mid -\frac{\nabla ^2}2\mid \phi _j> ~~,
\end{equation}
and for the non-local contribution we have
\begin{eqnarray}
E_{NL}^{l}&=&\sum_{s=1}^{\sigma}\sum_{m=-l}^l~
\alpha_{lm}^s~\sum_{ij}~\sum_{\bf k}\omega_{\bf k}~b_{ij}({\bf k}) \times
\nonumber \\
&\times& \sum_{I=1}^{N_s}~F_{Iilm}^*({\bf k})~F_{Ijlm}({\bf k}) ~~,
\label{enl}
\end{eqnarray}
where, in a plane waves basis set, $F_{Iilm}({\bf k})$ assumes the
following form:
\begin{equation}
F_{Ijlm}({\bf k})=\sum_{\bf g}~w_{lm}^s({\bf g+k})~
\exp\left(i{\bf g\cdot R}_I\right)~\phi_i({\bf g}) ~~,
\end{equation}
\noindent being $\{{\bf g}\}$ the reciprocal lattice vectors and ${\bf R}_I$
the coordinates of atom $I$ belonging to species $s$. The coefficient $w_{lm}^s$
is the projection of the ${\bf g}$-th plane wave component expanded around the
k-vector ${\bf k}$, onto the $lm$-th subspace according to the 
Kleinman-Bylander\cite{klei-byl} prescription:
\begin{equation}
w_{lm}^s({\bf g+k})=<\delta v_l^s \varphi_{lm}^s Y_{lm}\mid{\bf g+k}>\qquad ,
\end{equation}
with $\delta v_l^s$ the $l$-th angular momentum pseudopotential component
for species $s$, $\varphi_{lm}^s$ the corresponding pseudoatomic orbital, 
and $Y_{lm}$ the spherical harmonic functions. Equation (\ref{enl}) also 
contains a normalization factor, which is given by the expression 
{$\alpha_{lm}^s=<\delta v_l^s \varphi_{lm}^s \mid \varphi_{lm}^s>^{-1}$}. 
$N_s$ is the number of atoms of species $s$, and $\sigma$ is
the number of species described through angular-dependent pseudopotentials.

It is interesting to remark that, since now the factors $w_{lm}^s$ depend explicitly
on the k-vector, it is no longer possible to perform the BZ summation on the
coefficients $b_{ij}({\bf k})$ once and for all, as it is done for the 
kinetic and
local potential terms. Now this sum has to be carried out for each PW component
${\bf g}$, and this is likely to result in a costly computational scheme. So 
far we have not studied this issue in detail, because this method has only
been applied to hydrogen and sodium, where a local pseudopotential 
description is feasible,
but we hope that alternatives can be found so that
the non-local part does not become the bottleneck of the calculation.

The calculation of the density is a subtle issue, because expression
(\ref{rhokdp}) involves a double sum of matrix products over all states,
occupied and empty. Even if it has been suggested~\cite{payne1} that
this operation is computationally convenient, we have found it more efficient
to first transform the orbitals at $\Gamma $ by multiplying them with the 
rotation
matrices $a_{ij}({\bf k})$, and then carry out a single summation over
the occupied states. 

The last term in (\ref{forbkdp}) is easier to compute in reciprocal space,
where it assumes the following expression:
\begin{eqnarray}
\frac{\delta E_{NL}[\phi ,{\bf b}]}{\delta \phi _i^{*}({\bf g})}&=&
-\sum_{s=1}^{\sigma}\sum_{lm}\alpha_{lm}^s~\sum_{j}~
\sum_{\bf k}\omega_{\bf k}~b_{ij}({\bf k}) \times \\
&\times& w_{lm}^s({\bf g+k})
\sum_{I=1}^{N_s}~\exp\left(i{\bf g\cdot R}_I\right)~F_{Ijlm}({\bf k}) ~~.
\nonumber
\end{eqnarray} 

The forces on the nuclei are trivially unchanged from the original ones, as
long as local pseudopotentials are used. This is because the electronic
kinetic term does not depend explicitly on the nuclear positions. In the
case of angular-dependent pseudopotentials the contribution from the last
term in (\ref{enl}) to the force on atom $I$ belonging to species $s$, is
modified from the standard result in the following way:
\begin{eqnarray}
\frac{\partial E_{NL}}{\partial {\bf R}_I}&=&2
\sum_{lm}\alpha_{lm}^s~\sum_{ij}~\sum_{\bf k}\omega_{\bf k}~\times\nonumber \\
&\times&{\rm Re}\left( b_{ij}({\bf k})~\frac{\partial F_{Iilm}^*({\bf k})}
{\partial {\bf R}_I}~F_{Ijlm}({\bf k}) \right)\qquad .
\end{eqnarray}

The calculation of the stress matrix is also modified with respect to
standard calculations only for the kinetic ({\it kin}) and 
nonlocal pseudopotential ({\it NL}) contributions. 
The resulting expression for the kinetic contribution is:
\begin{eqnarray}
\sigma_{\alpha\beta}^{kin}&=&
\sum_{ij}B_{ij}^{(0)}~t_{ij\alpha\beta}^0+\sum_{ij}\left(
B_{ij\beta}^{(1)}~p_{ij\alpha}+B_{ij\alpha}^{(1)}~p_{ij\beta}
\right)+ \nonumber \\
&+& \sum_i B_{i\alpha\beta}^{(2)} ~~,
\label{strkdp}
\end{eqnarray}
\noindent with
\begin{equation}
t_{ij\alpha\beta}^0=<\phi _i\mid \frac{\partial^2}
{\partial x_\alpha \partial x_\beta}\mid \phi _j>
\end{equation}
\noindent and
\begin{eqnarray}
B_{ij\alpha}^{(1)}&=&\sum_{\bf k}\omega_{\bf k}~k_\alpha~b_{ij}({\bf k})
\\ \nonumber
B_{i\alpha\beta}^{(2)}&=&\sum_{\bf k}\omega_{\bf k}~
k_\alpha~k_\beta~b_{ii}({\bf k})\qquad .
\end{eqnarray}
The nonlocal contribution is instead:
\begin{eqnarray}
\sigma_{\alpha\beta}^{NL}&=&-2\sum_{s=1}^{\sigma}\sum_{lm}\alpha_{lm}^s~
\sum_{ij}~\sum_{\bf k}\omega_{\bf k}~b_{ij}({\bf k}) \times \nonumber \\
&\times& \sum_{I=1}^{N_s}
{\rm Re}\left(\frac{\partial F_{Iilm}^*({\bf k})}{\partial
{\bf h}_{\alpha\beta}}~F_{Ijlm}({\bf k})\right)\qquad ,
\label{stress_nl}
\end{eqnarray}
where ${\bf h}_{\alpha\beta}$ is the matrix of the three
primitive Bravais vectors~\cite{par-rah}, and the partial derivative in
(\ref{stress_nl}) is given in Ref.~\cite{nielsen}.

\end{document}